\documentclass[%
twocolumn,
amsmath,amssymb,
aip,
jcp
]{revtex4-1}

\usepackage{graphicx}
\usepackage{dcolumn}
\usepackage{bm}
\usepackage{hyperref}
\usepackage[english]{babel}
\usepackage{color}
\usepackage{epstopdf}

\begin{document}

\title{The effect of confinement on the solid-liquid transition in a core-softened potential system}

\author{Yu. D. Fomin}
\email{fomin314@mail.ru}
\author{E. N. Tsiok}
\author{V. N. Ryzhov}
\affiliation
{Institute for High Pressure Physics RAS, 108840 Kaluzhskoe shosse, 14, Troitsk, Moscow,
Russia}

\date{\today}

\begin{abstract}
We present a comparative computer simulation study of the phase
diagrams and anomalous behavior of two-dimensional ($2D$) and
quasi-two-dimensional ($q2D$) classical particles interacting with
each other through isotropic core-softened potential which is used
for a qualitative description of the anomalous behavior of water
and some other liquids. We have shown that at the low density part
of the phase diagram an increase in the width of the confining
slit pore leads to a considerable decrease in the melting
temperature while at high densities the melting temperature is
almost unchanged.
\end{abstract}


\maketitle

\section{Introduction}

Two-dimensional and quasi-two-dimensional systems that exist under
the conditions of strong spatial constraint in one direction
(confinement) are widespread in nature and technology. Processes
involving adsorbed ions, nano- and microparticles, as well as
colloidal suspensions and emulsions are ubiquitous in physical,
chemical and biological systems, and in technologies of new
materials \cite{r1,r2,r3,r4}. Confinement leads to the pronounced
role of long-range fluctuations in the physics of phase
transitions in two-dimensional ($2D$) systems, which transitions,
because of this, turn out to be even more diverse than $3D$
analogs.

Despite the abundance of publications over the past years, the
nature of $2D$ melting is one of the most intriguing problems in
condensed matter physics. In contrast to the $3D$ case, where
melting always occurs through a standard first order transition,
several scenarios are known for the microscopic description of
$2D$ melting. The main reason for this difference is a significant
increase in fluctuations in $2D$ systems compared to the case of
three dimensions. Peierls \cite{p1,p2}, Landau \cite{lan}, and
later Mermin~\cite{mermin} showed that in two dimensions,
long-range crystalline (translational) order could not exist due
to thermodynamic fluctuations and was transformed into a
quasi-long-range order, characterized by a slow power decrease in
correlations. On the other hand, the real long-range orientational
order (the orientations of the `bonds' between the nearest
particles) exists in two dimensions. At high temperatures, the
system turns into ordinary isotropic liquid.

The most popular theory of $2D$ melting (we will call it  the
first scenario) is the
Berezinsky-Kosterlitz-Tauless-Halperin-Nelson-Young theory (the
BKTHNY theory) \cite{b1,kosthoul73,halpnel1,halpnel2,halpnel3}
(see reviews \cite{kost,ufn1,ufn2,strRMP}). In accordance with the
BKTHNY theory, $2D$ melting occurs through two continuous
transitions with an intermediate hexatic phase. The transitions in
two-dimensional systems occur due to the formation of topological
defects. During the transition from crystal to  hexatic phase,
bound dislocation pairs dissociate at  certain temperature $T_m$,
turning  quasi-long-range translational order into  short-range
translational order, as well as  long-range orientational order
into  quasi-long-range orientational order. The new phase, which
has quasi-long-range orientational order and a zero shear modulus,
is called a hexatic phase. The resulting free dislocations can be
considered as bound disclination pairs that dissociate at certain
temperature $T_i$, and the system is transformed into  isotropic
liquid with short-range correlations. Both transitions are
continuous, of the Berezinskii-Kosterlitz-Thouless type, in
contrast to ordinary melting in 3D which is a first-order phase
transition. The BKTHNY theory has been confirmed in a number of
experiments, for example, in experiments with colloidal model
systems with repulsive magnetic dipole-dipole interaction
\cite{keim1,zahn,keim2,keim3,keim4}.

On the other hand, first order $2D$ melting can also occur (the
second scenario). For example, as it was shown in Ref.
\cite{chui83}, at low dislocation core energy  the dissociation of
bound dislocation pairs was preempted by the proliferation of
grain boundaries leading to first-order melting transition. A
similar scenario was discussed in
Refs.\cite{ryzhovTMP,ryzhovJETP}. Within the framework of the
density functional theory of crystallization in $2D$, the possible
dependence of the melting scenarios on the shape of the potentials
was studied in Refs. \cite{rto1,rto2,RT3,RT4}. A unified approach
to the description of first-order and
Berezinskii-Kosterlitz-Thouless  type melting with the help of
Landau's theory of phase transitions was proposed in Refs.
\cite{RT1,RT2}.

It should be noted that many early experiments and computer
simulations show contradictory results that reveal the following
trend: while the melting scenario of systems with long-range
potentials basically corresponds to the BKTHNY theory, for a long
time most researchers adhered to the view that systems with
short-range potentials and hard sphere potentials melted via a
first-order phase transition (see the discussions in reviews
~\cite{ufn1,ufn2,strRMP}).

Several years ago another (third) scenario of melting was proposed
~\cite{foh1,foh2,foh3,foh4,hsn}. In contrast to the BKTHNY theory,
computer simulations have shown that the melting of hard disk and
short-range potential systems can occur through two transitions:
the solid-to-hexatic phase transition occurs through the
continuous Berezinskii-Kosterlitz-Thouless transition and the
hexatic phase - isotropic liquid transition -- via a first-order
phase transition. One should note work ~\cite{foh4} (see also
\cite{physa2018,mp2019}), where the melting of a soft disk system
described by the potential of the form $1/r^n$ was considered. It
was shown that for $n<6$ the system melted in accordance with the
BKTHNY theory, and for $n>6$ - in accordance with the third
scenario.

In Refs.~\cite{foh7,hertz2018}, a $2D$ system with the Hertz
potential was discussed. The Hertz potential describes the elastic
energy of elastically deformed particles and can be used to
describe soft macromolecules, for example, micelles or star
polymers, as well as some colloidal systems. It was found that the
system under study could form a large number of ordered phases,
including a dodecagonal quasicrystal phase. In addition, the
melting scenarios of this system were analyzed. It was shown that,
depending on the position in the phase diagram, the system could
melt not only in accordance with the BKTHNY scenario and
first-order phase transition, but also in accordance with the
third scenario with one transition of the first order and one
continuous transition. Tricritical points at which the melting
scenario of the system changes, and a water-like density anomaly
(the thermal expansion coefficient is negative) were found
 \cite{hertz2018}.

When discussing experimental realizations of two-dimensional
systems it is necessary to keep in mind that real systems are
quasi-two dimensional, in which out-of-plane molecular motion
cannot be eliminated. The question to be answered is whether small
amplitude out-of-plane particle motion only produces small
quantitative corrections to the phase diagram predicted under the
assumption that particle motion is strictly two-dimensional, or it
generates qualitative changes to the phase diagram and transition
scenarios. For instance, in Ref. \cite{rice1} Zangi and Rice
presented the results of extensive simulations of several phase
transitions in a quasi-two-dimensional system with the Marcus-Rice
(MR) potential \cite{rice3}. They found first-order
liquid-to-hexatic and hexatic-to-solid transitions, in agreement
with the experimental results of Marcus and Rice \cite{rice3}. The
results of the simulations also reveal an isostructural
solid-to-solid transition. Frydel and Rice \cite{rice2} have
compared the phase diagrams of $q2D$ and $2D$ systems composed of
particles with the MR interaction. Both systems undergo first
order solid I -- solid II and solid II -- solid III isostructural
transitions. The introduction of out-of-plane motion shifts the
low-density portion of the phase boundaries to lower temperatures.
The liquid - solid I coexistence line is nearly the same for the
two systems. The solid II -- solid III transition is shifted to
lower temperature and to higher density in the $q2D$ system.
Frydel and Rice \cite{rice2} suggested that the change from $2D$
to $q2D$ confinement had non-negligible effect on the nature of
phase transitions in the MR system.

On the other hand, in Ref.~\cite{foh3} it was shown that the third
melting scenario observed for a real $2D$ hard disk system
\cite{foh1} persisted for a quasi-$2D$ hard sphere system with
deviation of particle motions from the plane to a distance of up
to half the particle diameter.

It is well known that confined systems can have many unusual
structures, which are not observed in ordinary $3D$ substances.
One of the most important examples is the discovery of square ice
which is formed when water is confined between graphene planes
\cite{geim}. Many other complex structures were predicted to be
formed when water is confined in nanotubes \cite{nt1,nt2,nt3,nt4}.

Even simple systems like noble gases modeled by the Lennard-Jones
(LJ) potential demonstrate complex behavior under confinement
\cite{ljslit1,ljslit2,ljslit3,ljslit4}. When the LJ system is
placed into a slit pore it demonstrates splitting into a different
number of layers. Moreover, examining the two-dimensional ($2D$)
structure of these layers showed that their structure changed with
the pore width. As the width of the pore increases the structure
changes from $n$ layers with a square structure to $n$ layers with
a triangular structure, $n+1$ square layers, etc.

The main goal of the present paper is to compare the behavior of a
purely $2D$ system with the core-softened potential and a one
layer quasi-two-dimensional system ($q2D$) composed of the same
particles and confined in slit pores. This comparison can give
some qualitative hints for understanding the role of confinement
in the behavior of  systems with core-softened potentials with two
length scales, which are used for a qualitative description of
water-like anomalous liquid behavior.

\begin{figure}
\includegraphics[width=8cm]{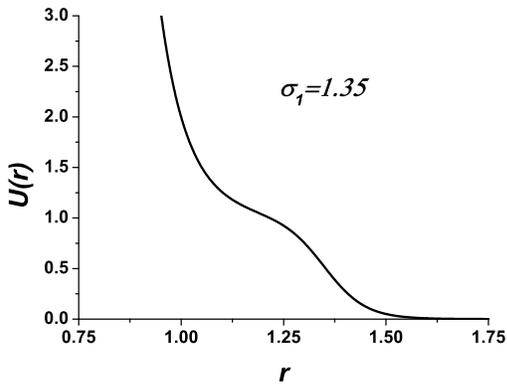}%

\caption{\label{fig1}  The potential (\ref{pot}) with $\sigma =
1.35$.}
\end{figure}

In the present paper we consider a core-softened system which is
characterized by the interaction potential \cite{jcp2008,pre2009}
(see Fig.~\ref{fig1}):
\begin{equation}\label{pot}
  U(r)/\varepsilon= \left( \frac{d}{r} \right)^n+0.5 \left( 1 - \tanh(k(r- \sigma))
  \right),
\end{equation}
where $n=14$, $k=10$ and parameter $\sigma /d=1.35$ determines the
width of the potential's repulsive shoulder. Below we express all
quantities in reduced units related to the potential parameters,
i.e. $\varepsilon$ and $d$ are used as energy and length units.

The properties of the $2D$ system described by the potential
(\ref{pot}) were extensively studied in a series of papers
\cite{we1,we2,we3,we4,we5}. The phase diagram of this system
contains several ordered structures: two triangular phases (with
low and high density), a square crystal and a dodecagonal
quasicrystal. The triangular phase with low density is of special
importance. This phase demonstrates reentrant melting behavior,
and the melting scenarios at $\rho < \rho_{max}$ (the left branch)
and $\rho > \rho_{max}$ (the right branch) ($\rho_{max}$ is the
density of the  melting line maximum) are different: at the left
branch a first-order transition takes place, while at the right
branch melting occurs in accordance with the third scenario. The
water-like anomalies such as those of density, structure and
diffusion were also found.

The water-like anomalies were found in a $q2D$ core softened
system confined between two fixed hydrophobic plates
\cite{barbosa}. The number of layers (two or three) depended on
the range of confining distances. In the case of three layers
crystallization occurred in the contact layers, the middle layer
stayed liquid. Also the temperatures of density and diffusion
maxima decreased compared with the bulk values. Although the
authors did not calculate the melting temperature one can suppose
that it becomes lower with pore width expansion because the
temperature of a density maximum and  of diffusion coefficient
maximum  and  minimum decreases with an increase in the pore
width.

In the present paper we consider a $q2D$ system with the
interaction potential (\ref{pot}) in slit pores of different
width. We have shown that the pore width strongly affects the low
density triangular phase while the high density part of the phase
diagram remains nearly unaffected.

\section{System and methods}

We simulated a system of 20000 particles in a slit pore by means
of a molecular dynamics method. An $NVT$ - constant ensemble and
the Nose-Hoover thermostat were used in order to keep the
temperature fixed. Periodic boundary conditions were maintained
along the x and y axes. Two structureless walls parallel to the XY
plane were placed at $z=0$ and $z=H$. The width of pore $H$ varied
from $H/ \sigma=0.3$ up to $H/ \sigma =3.0$. The interaction of a
wall with particles is described by the Lennard-Jones 9-3
potential (see Fig. 2):

\begin{equation}\label{potpore}
U_{pore}/ \varepsilon=  \left( \frac{\sigma}{r} \right)^9 - \left(
\frac{\sigma}{r} \right)^3.
\end{equation}

\begin{figure}
\begin{center}

\includegraphics[width=8cm]{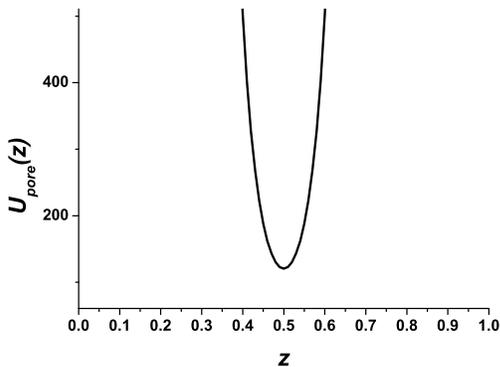}%
\end{center}

\caption{\label{fig:fig2}  Potential $U_{pore}(r)$ of
particle-wall interaction from equation (\ref{potpore})}
\end{figure}

The system was equilibrated for $2 \cdot 10^7$ time steps, with a
subsequent $3 \cdot 10^7$ simulation run. The time step was set to
$dt=0.001$.  In the present paper we concentrate on the case where
only one layer of particles is formed. It takes place when the
pore width is $H \leq 1.8$. The case of $H>1.8$ will be considered
in a further publication. Therefore we introduced a
two-dimensional number density of the system: $\rho= N/A$, where A
is the square of the system. All densities below mean this planar
density. The density was varied from $\rho_{min}=0.4$ up to
$\rho_{max}=1.24$.

We have found that the system can form several different
crystalline phases - triangular and square ones and a dodecagonal
quasicrystal. We investigated the limits of stability of these
phases and the melting scenarios using the method employed in our
previous publications. This method is based on a combination of
the data from the equation of state, in particular, the presence
or absence of the Mayer-Wood loop, the diffraction patterns, the
radial distribution functions, the orientational and translational
order parameters and their correlation functions.

The diffraction pattern is calculated as $S(\bf{k})= < \frac{1}{N}
\left(\sum_i^Ncos(\bf{kr}_i)\right)^2+\left(\sum_i^Nsin(\bf{kr}_i)\right)^2>$.
In the case of an ordered structure the diffraction pattern shows
clear peaks of intensity, while in liquid phase it demonstrates a
uniform distribution of intensities.


One can define the orientational order parameter (OOP) of a
triangular lattice as \cite{halpnel1,halpnel2,ufn1}

\begin{equation}
\Psi_6({\bf r_i})=\frac{1}{n(i)}\sum_{j=1}^{n(i)} e^{i
n\theta_{ij}}\label{psi6loc},
\end{equation}
where $\theta_{ij}$ is the angle of the vector between particles
$i$ and $j$ with respect to a reference axis and the sum over $j$
is counting the $n(i)$ nearest neighbors of $j$. The nearest
neighbors are defined by the Voronoi tesselation. The average OOP
over the whole system gives global orientational order:

\begin{equation}
\psi_6=\frac{1}{N}\left<\left|\sum_i \Psi_6({\bf
r}_i)\right|\right>.\label{psi6}
\end{equation}

The translational order parameter (TOP) is defined as
\cite{halpnel1,halpnel2,ufn1}:
\begin{equation}
\psi_T=\frac{1}{N}\left<\left|\sum_i e^{i{\bf G
r}_i}\right|\right>, \label{psit}
\end{equation}
where ${\bf r}_i$ is the position vector of particle $i$ and {\bf
G} is the reciprocal-lattice vector of the first shell of the
crystal lattice.

The correlation functions of OOP and TOP which contain information
on long-range or quasi-long-range ordering in the system were also
calculated in the present work. Orientational correlation function
(OCF) $G_6(r)$ is given by the following expression:
\begin{equation}
G_6(r)=\frac{\left<\Psi_6({\bf r})\Psi_6^*({\bf 0})\right>}{g(r)},
\label{g6}
\end{equation}
where $g(r)=<\delta({\bf r}_i)\delta({\bf r}_j)>$  is a radial
distribution function. In the hexatic phase the long range
behavior of $G_6(r)$ has the form $G_6(r)\propto r^{-\eta_6}$ with
$\eta_6 \leq \frac{1}{4}$ \cite{halpnel1, halpnel2}, while in
isotropic liquid it decays exponentially. Since $2D$ crystals are
characterized by long-range orientational order, OCF has flat
shape in crystalline phase.

The translational correlation function (TCF) is calculated as
\begin{equation}
G_T(r)=\frac{<\exp(i{\bf G}({\bf r}_i-{\bf r}_j))>}{g(r)},
\label{GT}
\end{equation}
where $r=|{\bf r}_i-{\bf r}_j|$. Since $2D$ crystals demonstrate
quasi long-range translational order TCF decays algebraically in
crystalline phase: $G_T(r)\propto r^{-\eta_T}$ with $\eta_T \leq
\frac{1}{3}$ \cite{halpnel1, halpnel2}. In the hexatic phase and
isotropic liquid $G_T$ decays exponentially.


\section{Results and Discussion}

\begin{figure}
\includegraphics[width=6cm,height=4cm]{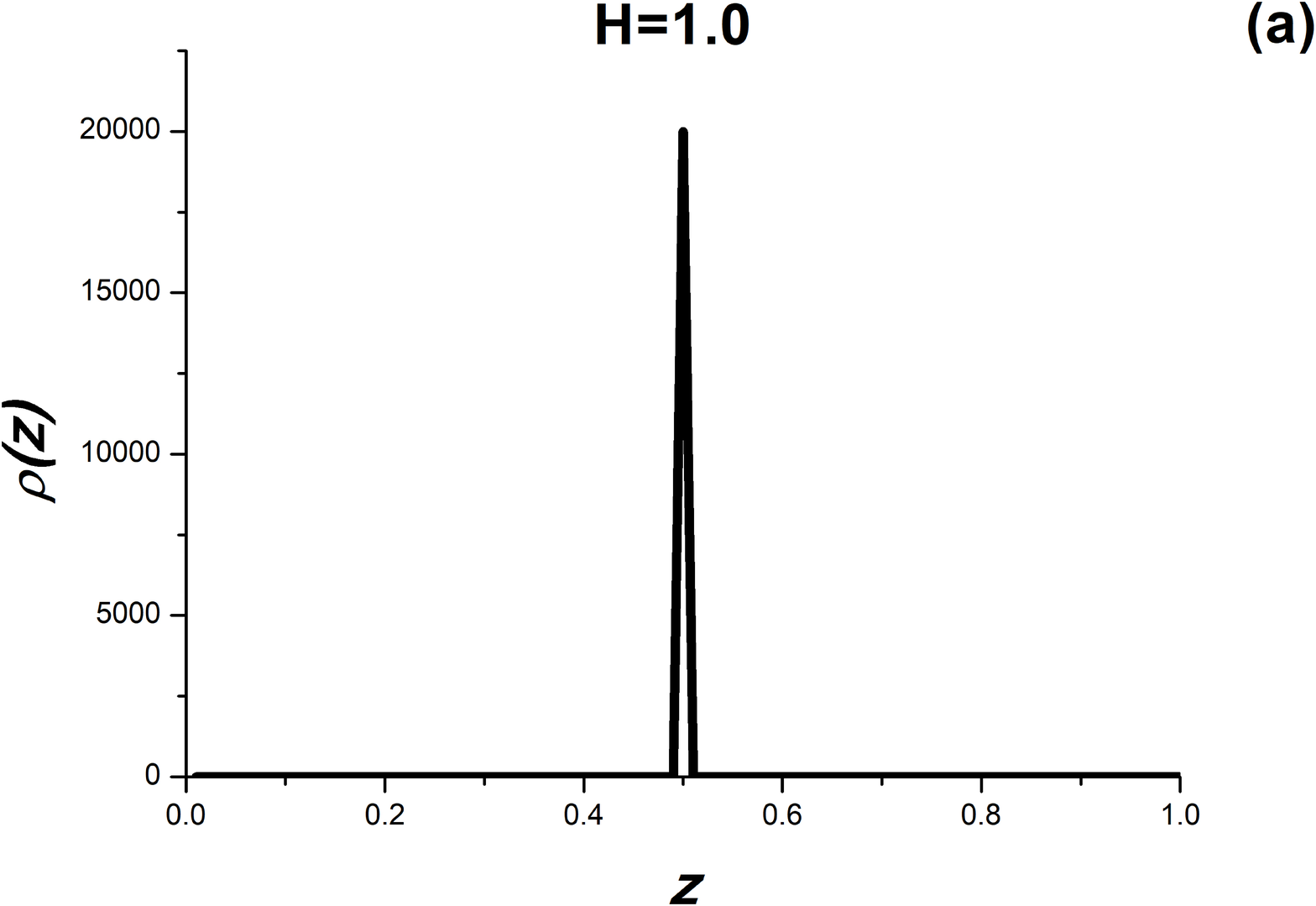}%

\includegraphics[width=6cm,height=4cm]{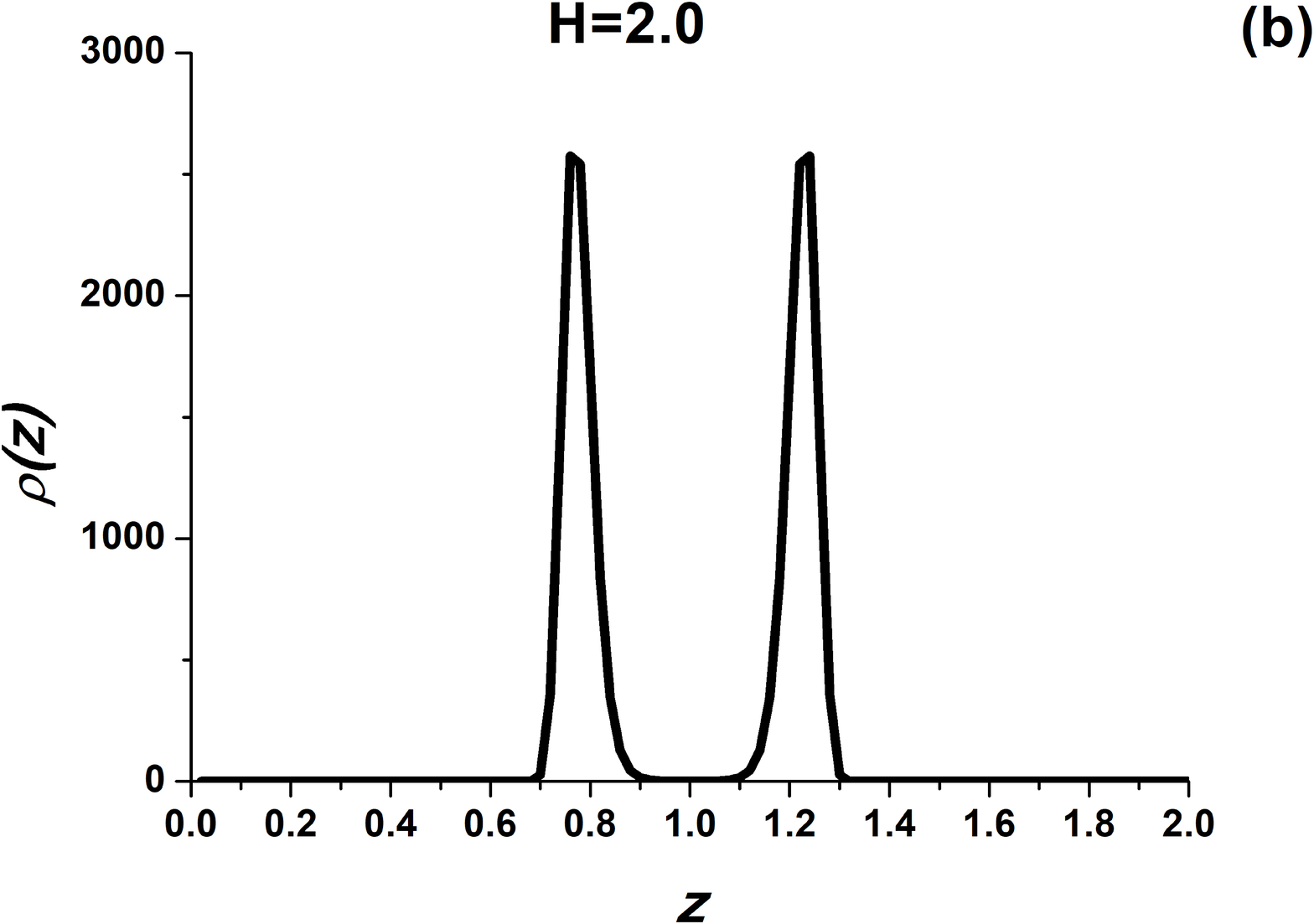}%

\caption{\label{zdistr} The density distribution at $H=1$ (a) and
$H=2$ (b).}
\end{figure}

First of all we investigated the distribution of density along the
z axis at high two-dimensional density $\rho=1.05$. Obviously in
the case of small H the system formed a single layer and we
intended to find out at what H the system would split into two
layers. Figs. \ref{zdistr} (a) and (b) show density distribution
at several values of $H$ ($H=1$, $H=2$). Our investigation has
shown that at $H \leq 1.8$ the distribution of density
demonstrated one peak while at $H>1.8$ two peaks were detected,
i.e., two layers were formed. Below we will discuss only systems
with a single layer, i.e. we restrict ourselves to $H \leq 1.8$.
We have also found that the behavior of all systems with $H\leq
1.8$ was qualitatively similar. Because of this most of the
results are demonstrated for the system with $H=1.0$.

\begin{figure}
\includegraphics[width=8cm]{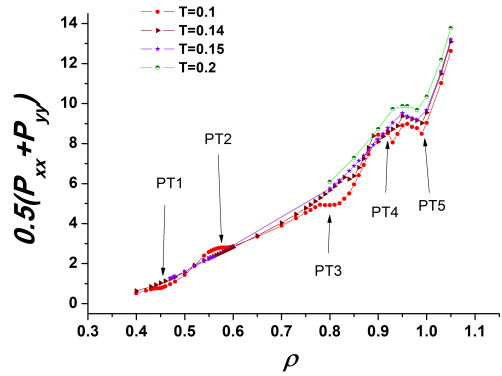}%

\caption{\label{allt01h1} Isotherms of the system in the pore with
$H=1$. Labels PT1,...,PT5 show the peculiarities of the equation
of state which signalize the appearance of phase transitions.}
\end{figure}

Fig. \ref{allt01h1} shows several isotherms of the system with
$H=1.0$. One can observe several peculiarities along these
isotherms, which are marked with labels PT1, ..., PT5. Such
peculiarities mean that phase transitions are highly likely at
these densities. Below we study these phase transitions and
identify all phases present in the system.

\begin{figure}
\includegraphics[width=8cm]{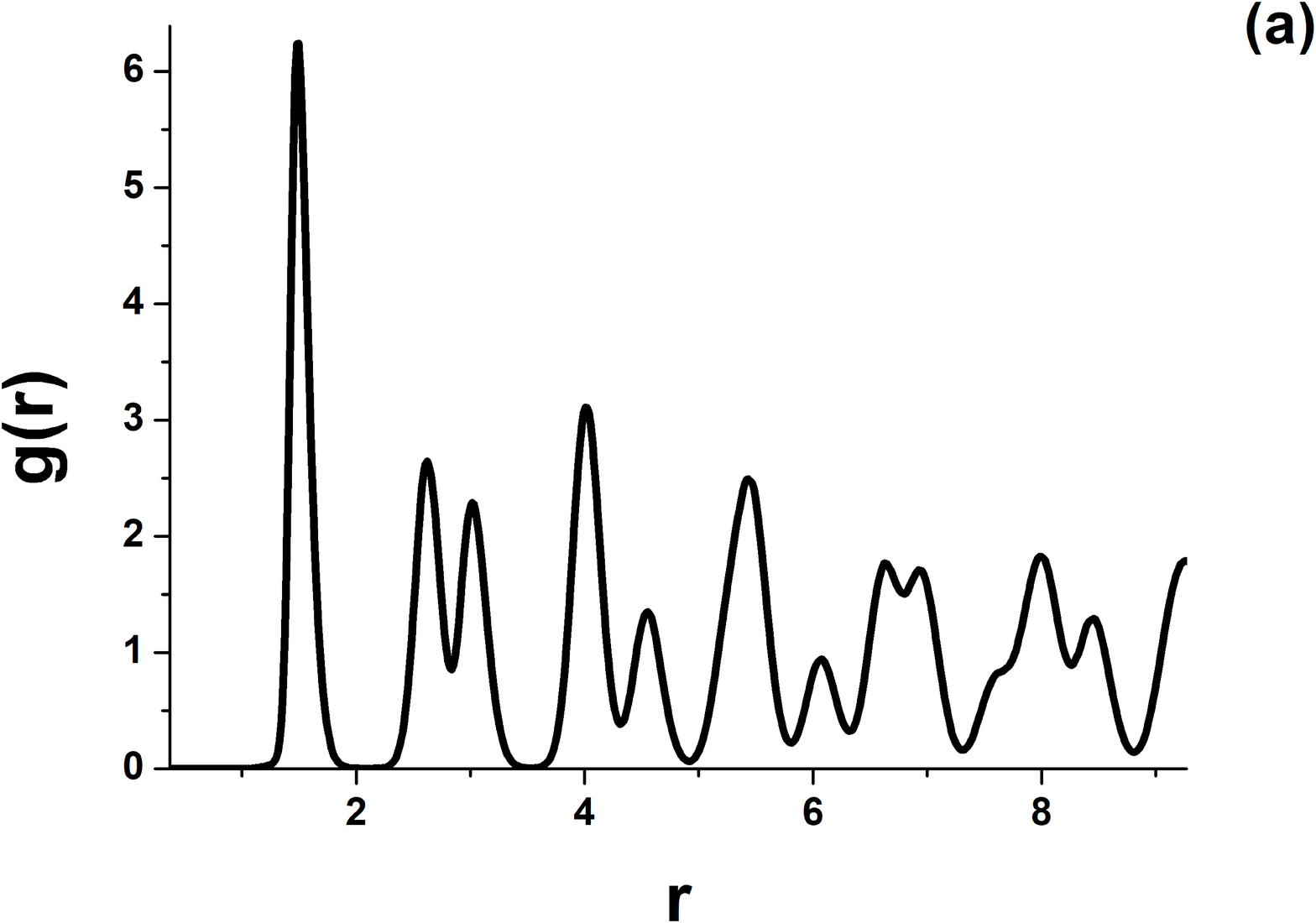}%

\includegraphics[width=8cm]{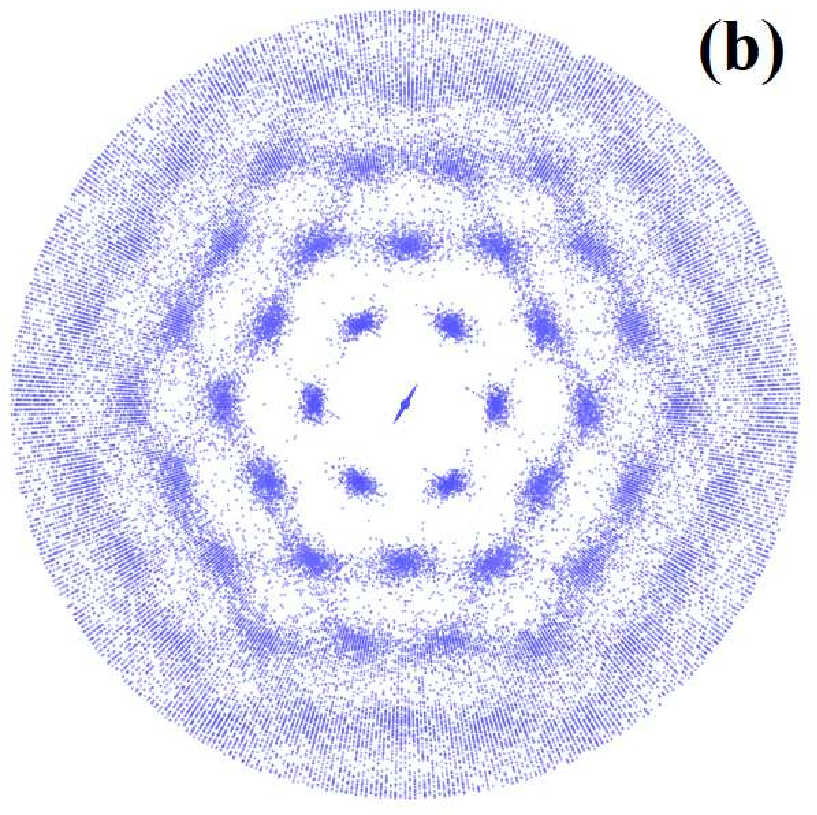}%

\caption{\label{rho05} (a) The radial distribution function of the
system with $H=1.0$ at $\rho=0.5$ and $T=0.1$; (b) the diffraction
pattern of the same system.}
\end{figure}

The first phase transition (PT1) is crystallization of low-density
liquid into a triangular crystal. Figs. \ref{rho05} (a) and (b)
show the rdf and the diffraction pattern of the system at $T=0.1$
and $\rho=0.5$. From these figures one can clearly see that the
system is in triangular phase at this point.

\begin{figure}
\includegraphics[width=8cm]{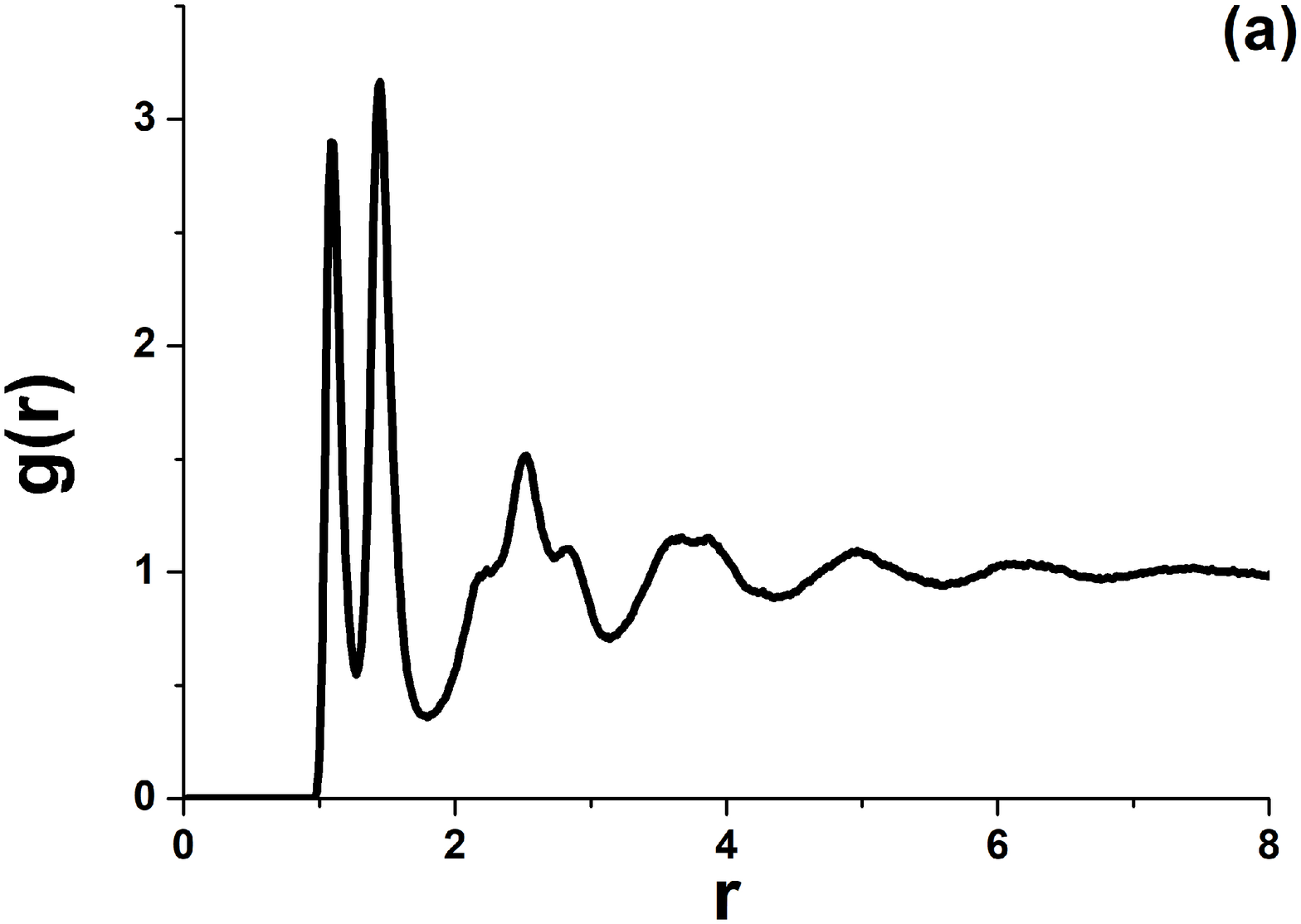}%

\includegraphics[width=8cm]{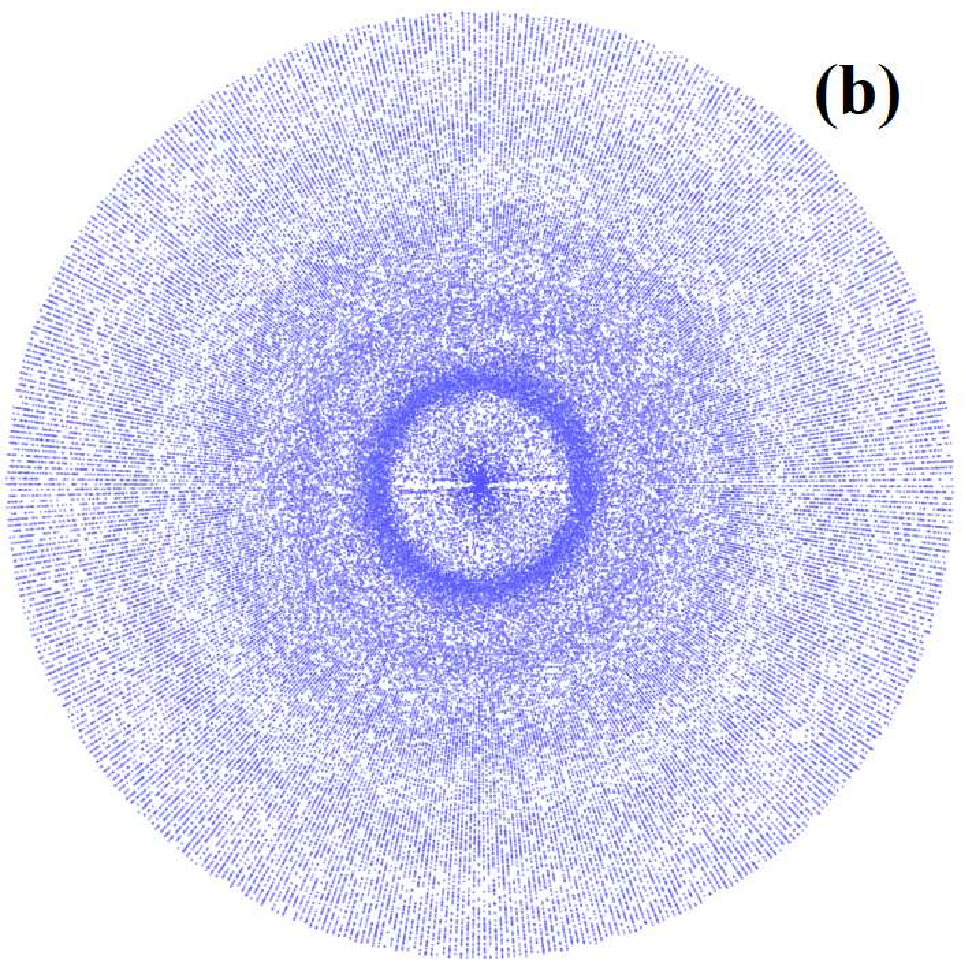}%

\caption{\label{rho065} (a) The radial distribution function of
the system with $H=1.0$ at $\rho=0.65$ and $T=0.1$; (b) the
diffraction pattern of the same system.}
\end{figure}

Figs. \ref{rho065} (a) and (b) show the rdf of the system at
$T=0.1$ and $\rho=0.65$.  The system is in a liquid state at this
point. It means that reentrant melting takes place in the system.

\begin{figure}
\includegraphics[width=8cm]{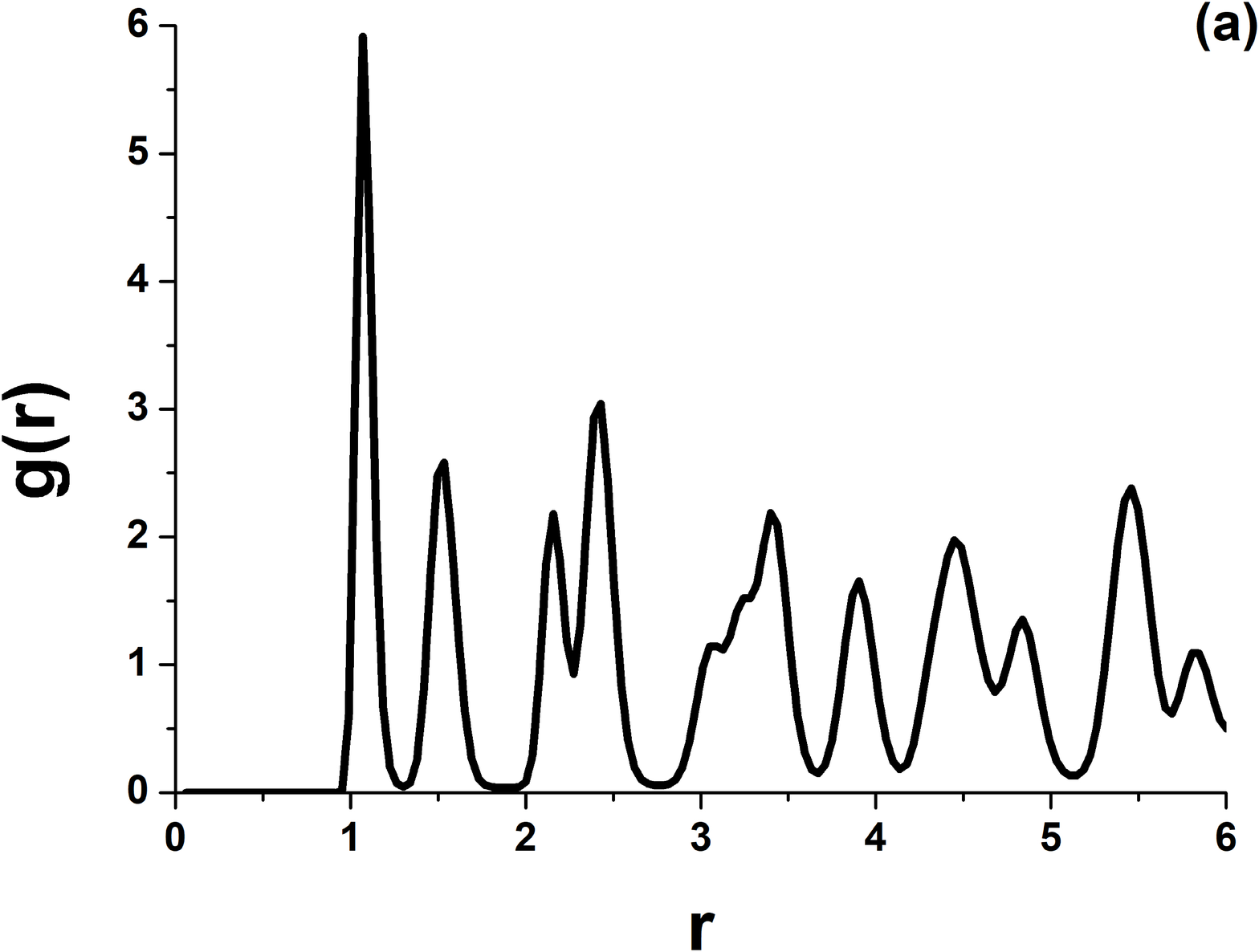}%

\includegraphics[width=8cm]{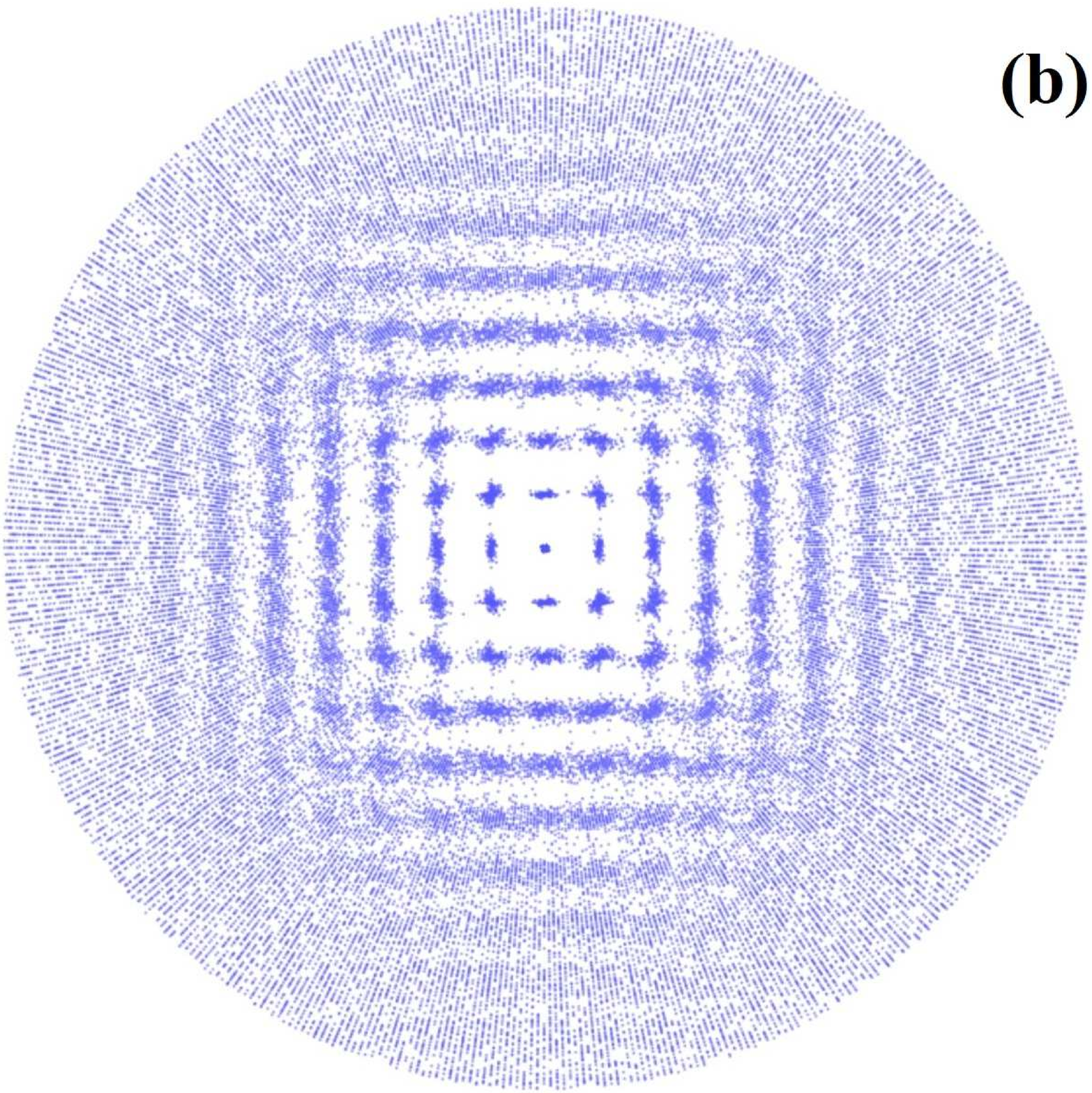}%

\caption{\label{rho085} (a) The radial distribution function of
the system with $H=1.0$ at $\rho=0.85$ and $T=0.1$; (b) the
diffraction pattern of the same system.}
\end{figure}

\begin{figure}
\includegraphics[width=8cm]{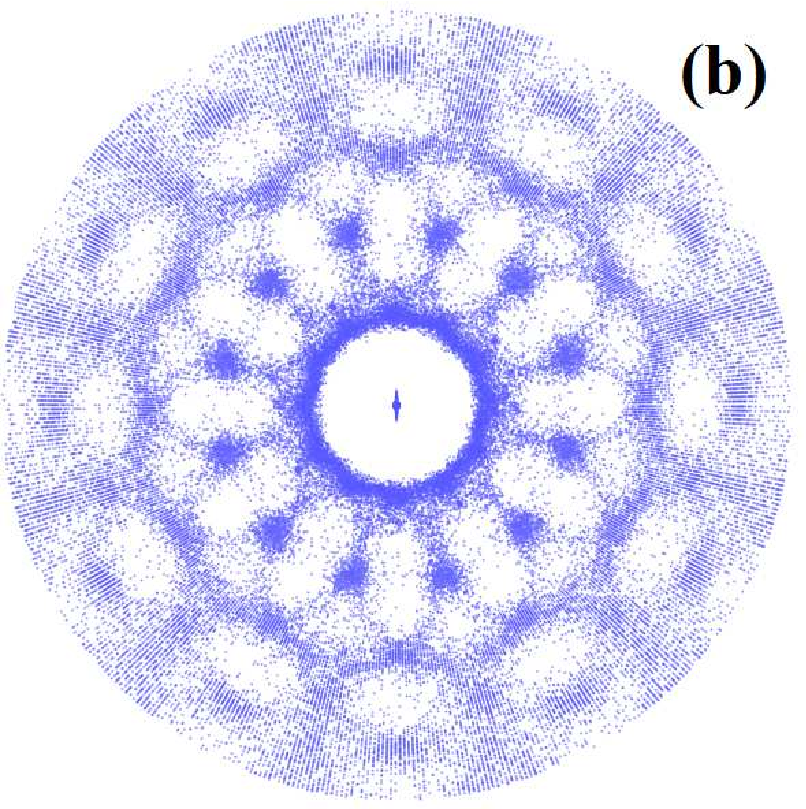}%

\includegraphics[width=8cm]{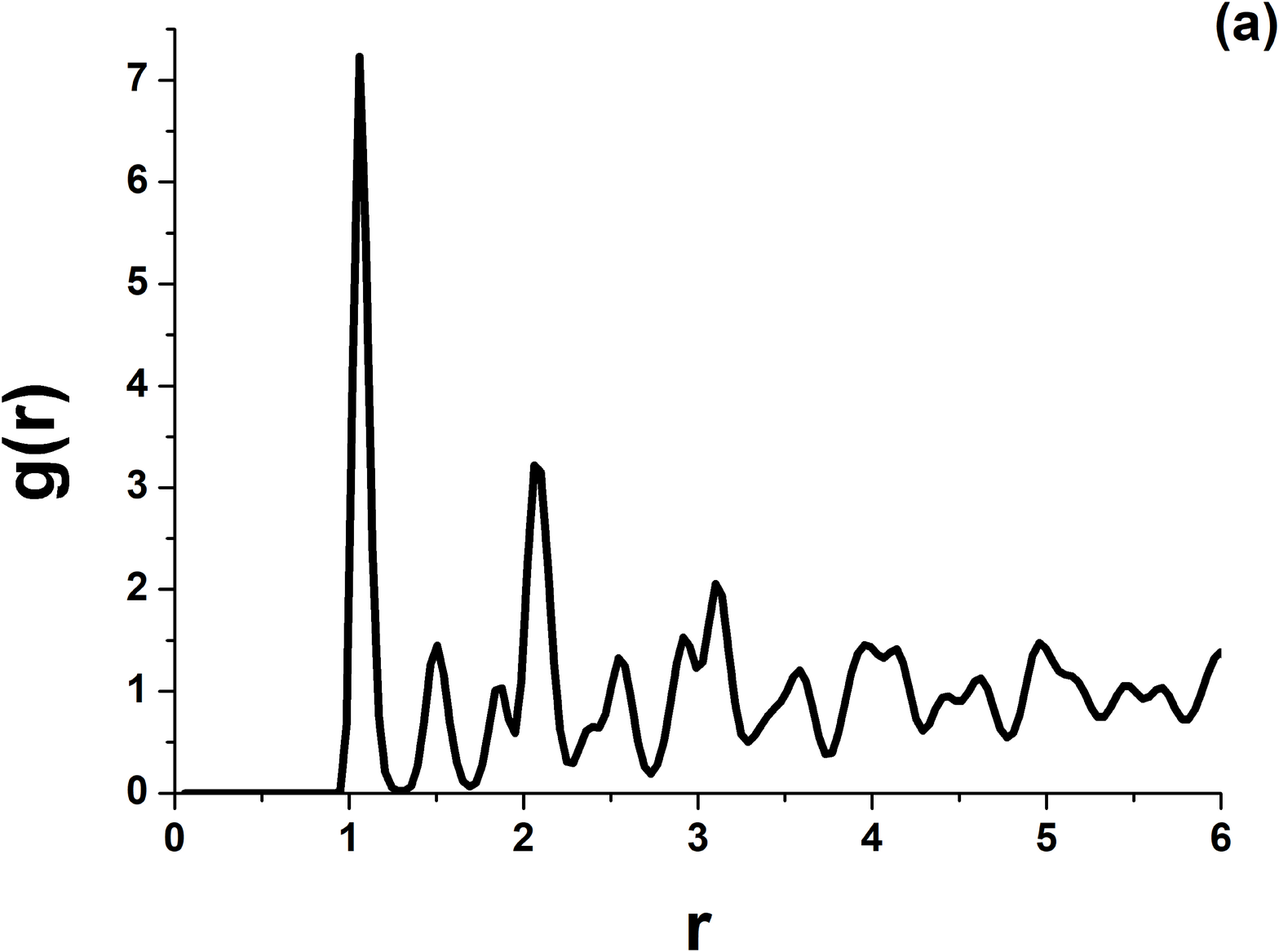}%

\caption{\label{rho093} (a) The radial distribution function of
the system with $H=1.0$ at $\rho=0.93$ and $T=0.1$; (b) the
diffraction pattern of the same system.}
\end{figure}

\begin{figure}
\includegraphics[width=8cm]{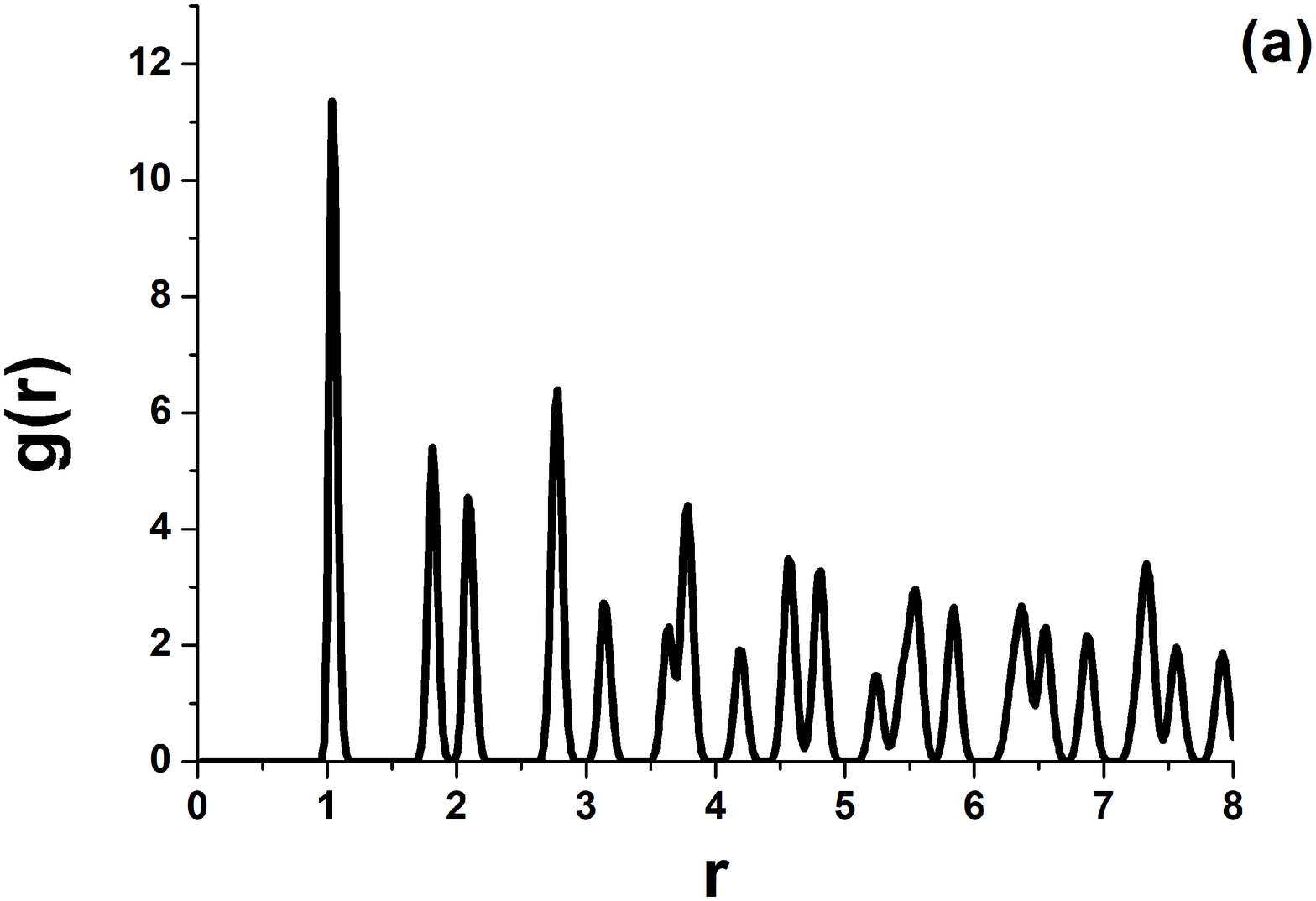}%

\includegraphics[width=8cm]{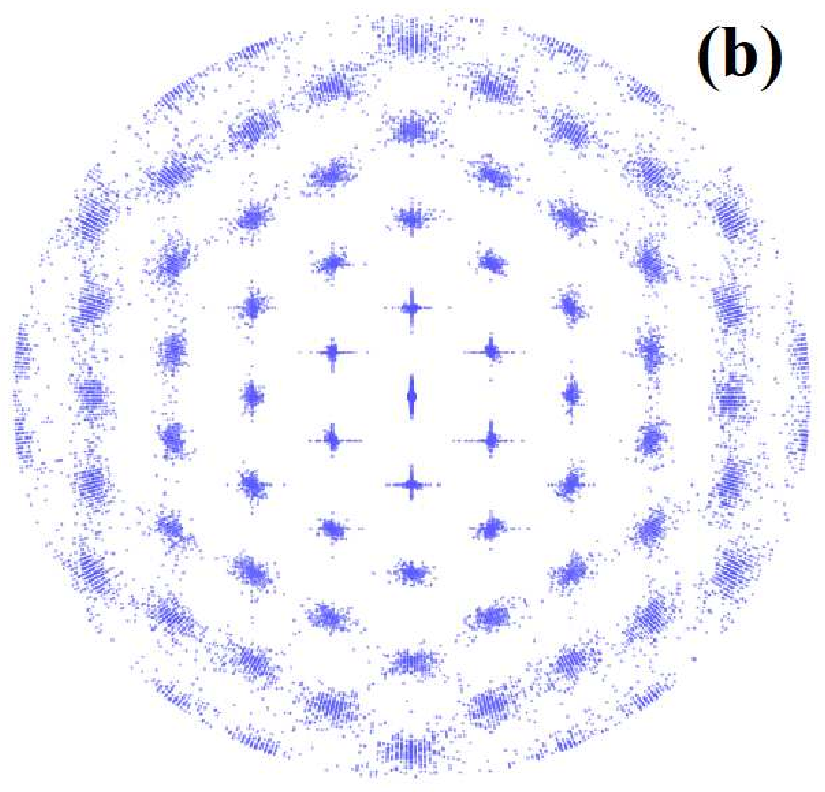}%

\caption{\label{rho105} (a) The radial distribution function of
the system with $H=1.0$ at $\rho=1.05$ and $T=0.1$; (b) the
diffraction pattern of the same system.}
\end{figure}

Further densification of the system leads to the appearance of a
square crystal (PT3). Figs. \ref{rho085} (a) and (b) show the rdfs
and diffraction patterns at $\rho=0.85$. The next phase transition
transforms the square crystal into a dodecagonal quasicrystal (see
Fig. \ref{rho093} for the rdfs and diffraction patterns). Finally
at high densities the system transforms into a triangular crystal
(see Fig. \ref{rho105} for the rdfs and diffraction patterns).
This sequence of phases is analogous to the purely two-dimensional
system studied in our previous publications
\cite{we1,we2,we3,we4,we5}.

Having identified all structures in the system we studied their
limits of stability and melting scenarios. First we considered the
low density triangular phase and its melting. Similar to the case
of the purely $2D$ system this phase demonstrated a maximum on the
melting line. Below we will call the branch of the melting line
with $\rho < \rho_{max}$ ($\rho_{max}$ is the density of the
melting line maximum) a left branch and the one with $\rho >
\rho_{max}$ -- a right branch. As we discussed in the
Introduction, in the case of the purely $2D$ system the melting
scenarios for the left and right branches are different: at the
left branch a first-order transition takes place, while at the
right branch melting occurs in accordance with the third scenario.
The water-like anomalies were also found.

\begin{figure}
\includegraphics[width=8cm]{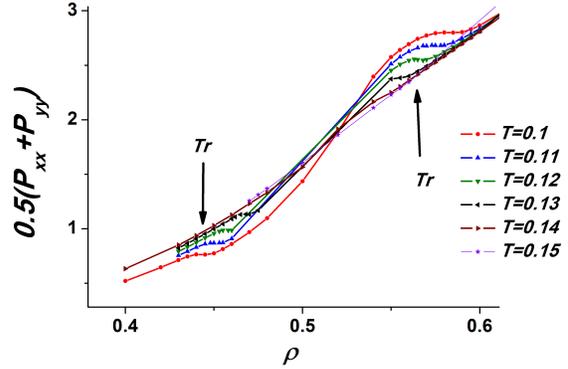}%

\caption{\label{eos-ld} The equation of state along several
isotherms crossing the low density triangular phase ($Tr$).}
\end{figure}

Fig. \ref{eos-ld} shows the equation of state along several
isotherms crossing the stability region of the low density
triangular crystal. One can see that two sets of loops appear on
these isotherms, which correspond to the crystallization of low
density liquid and reentrant melting of the crystal. Since the
Mayer-Wood loops are present in the system, both branches
demonstrate first order phase transition, i.e. melting should be
either first order transition, or the third scenario. In order to
determine the transition scenario we consider the correlation
functions of the orientational and translational order parameters.

\begin{figure}
\includegraphics[width=8cm]{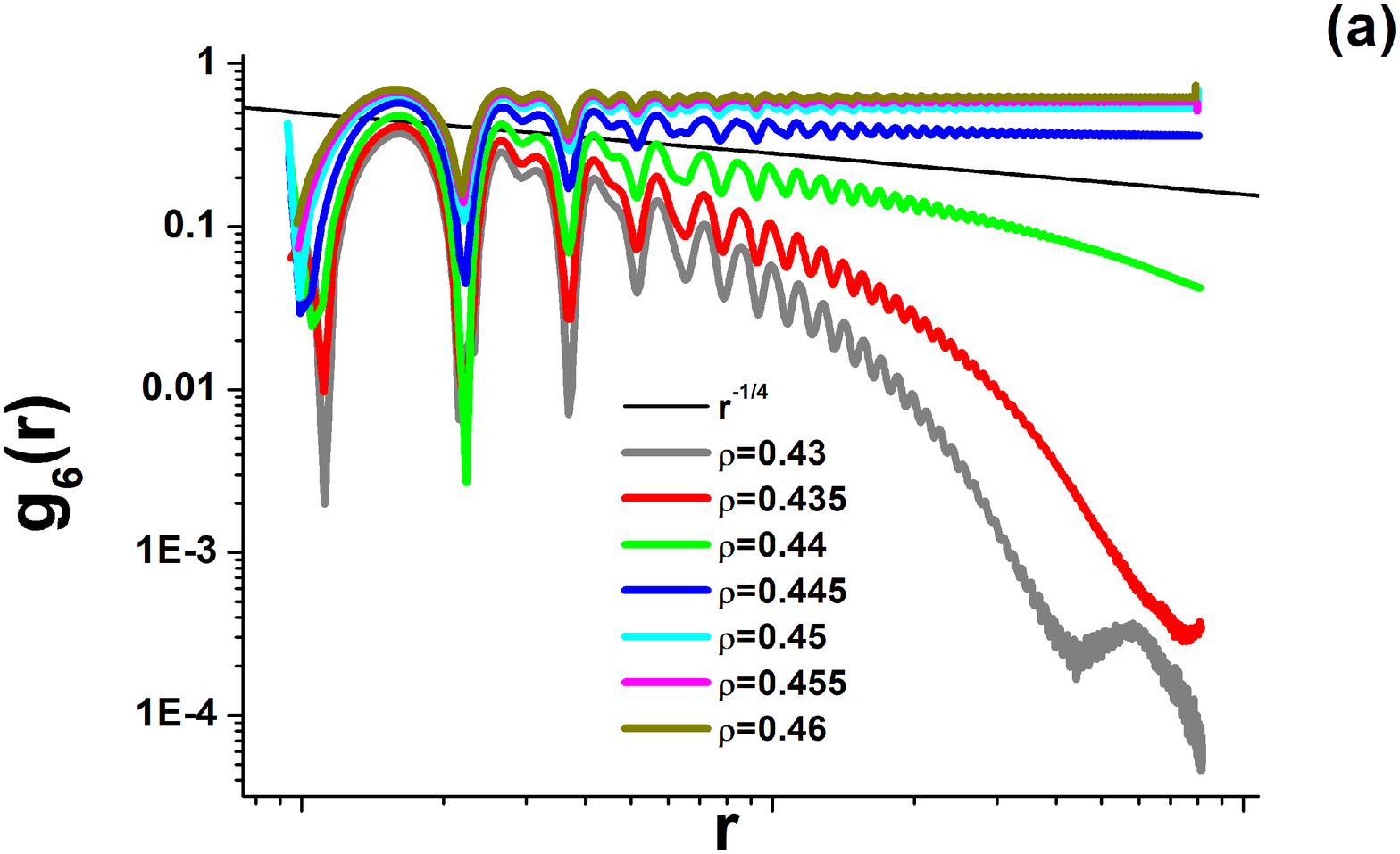}%

\includegraphics[width=8cm]{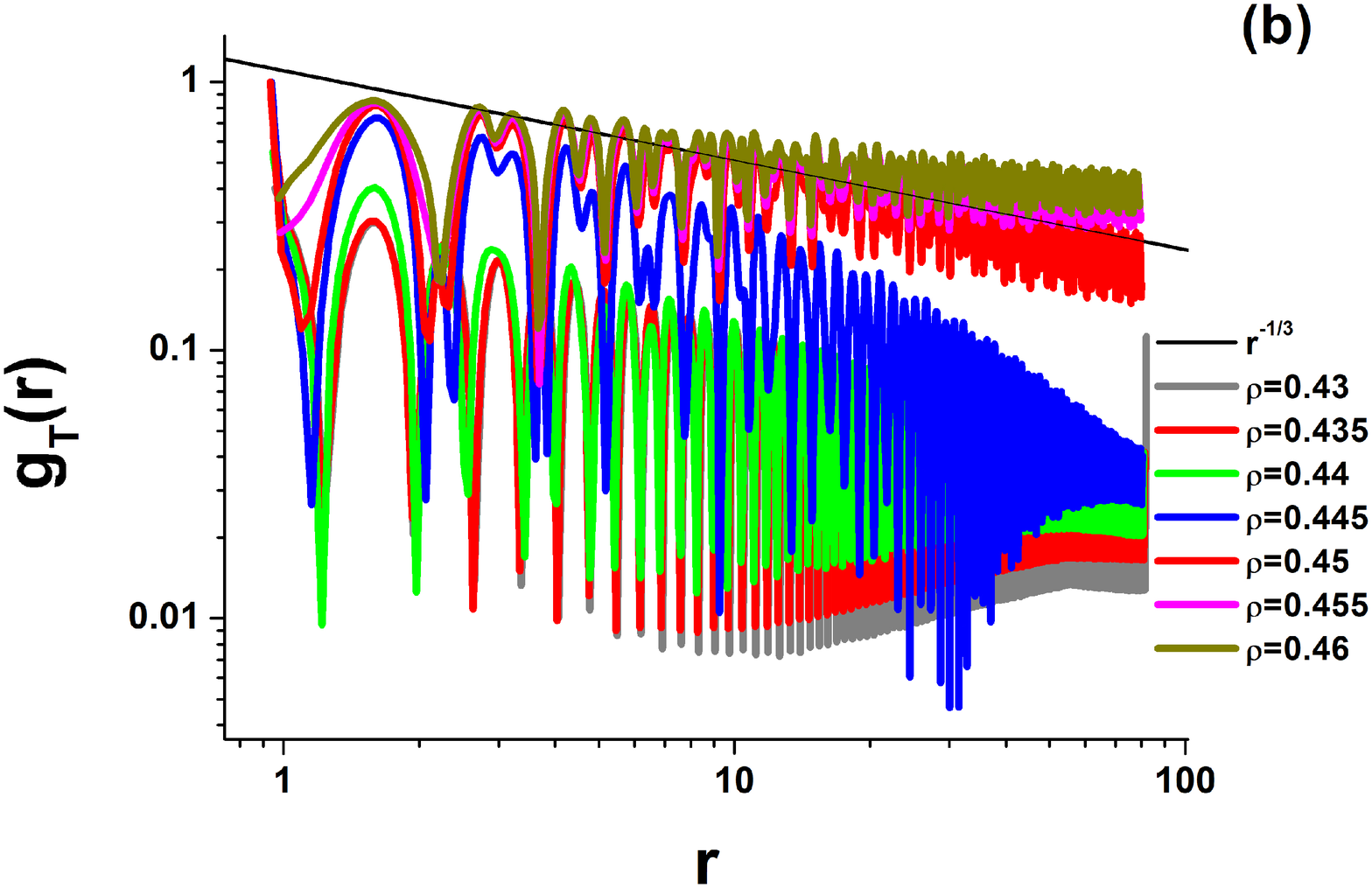}%

\caption{\label{g1-ld} (a) The orientational correlation functions
of the system with $H=1.0$ at the left branch of the melting line
of the low density triangular crystal at $T=0.1$. (b) The same for
the translational correlation functions.}
\end{figure}

Figs. \ref{g1-ld} (a) and (b) show the correlation functions of
OOP (a) and TOP (b) at $T=0.1$ crossing the left branch of the
melting line. Comparing the densities of the Mayer-Wood loop
existence region with the stability limits of the crystal and
hexatic phase obtained from correlation function behavior we
conclude that at the left branch of the low density phase diagram
melting occurs through a single first order phase transition
because both stability limits are located within the Mayer-Wood
loop existence region.

\begin{figure}
\includegraphics[width=8cm]{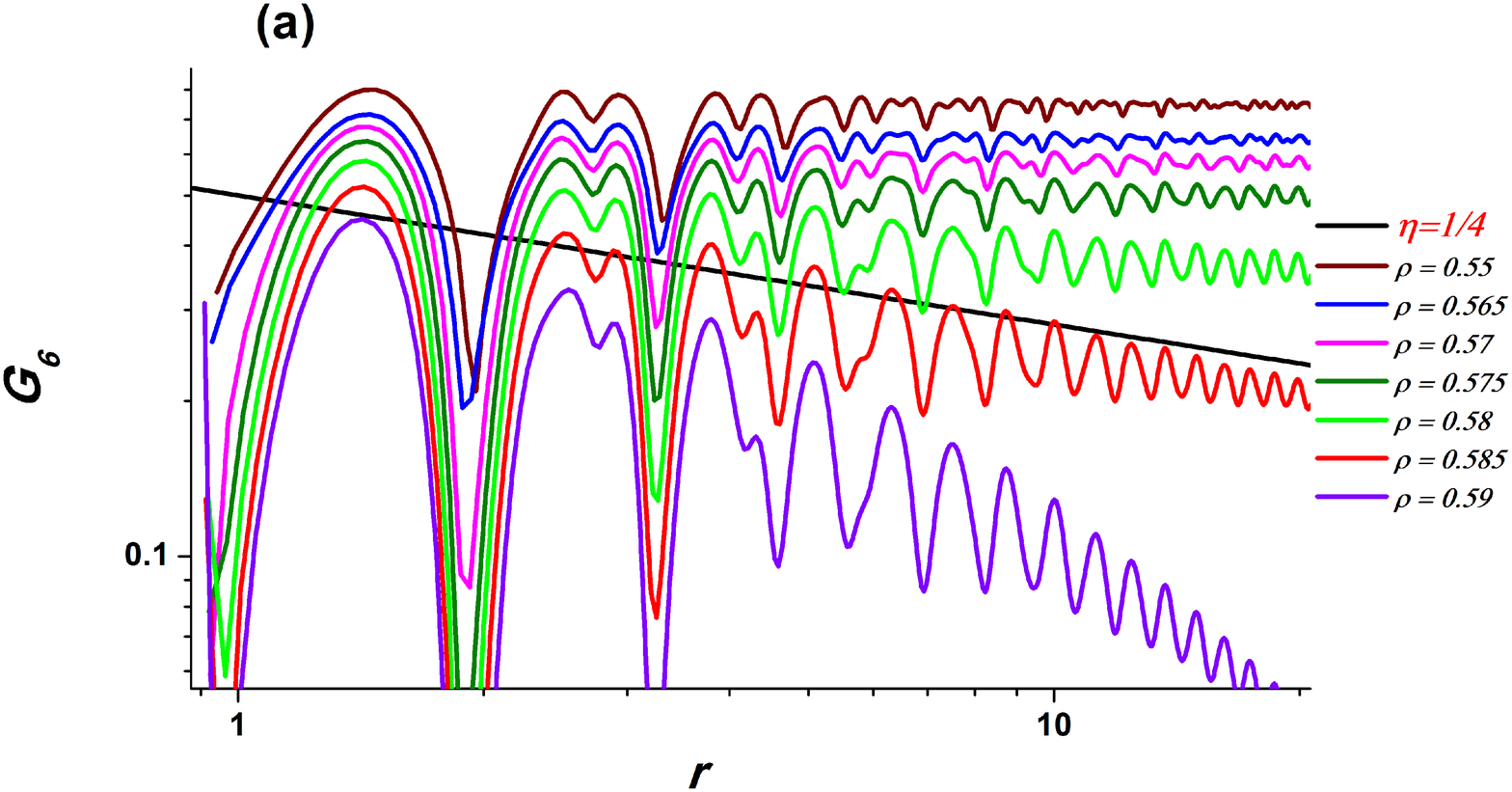}%

\includegraphics[width=8cm]{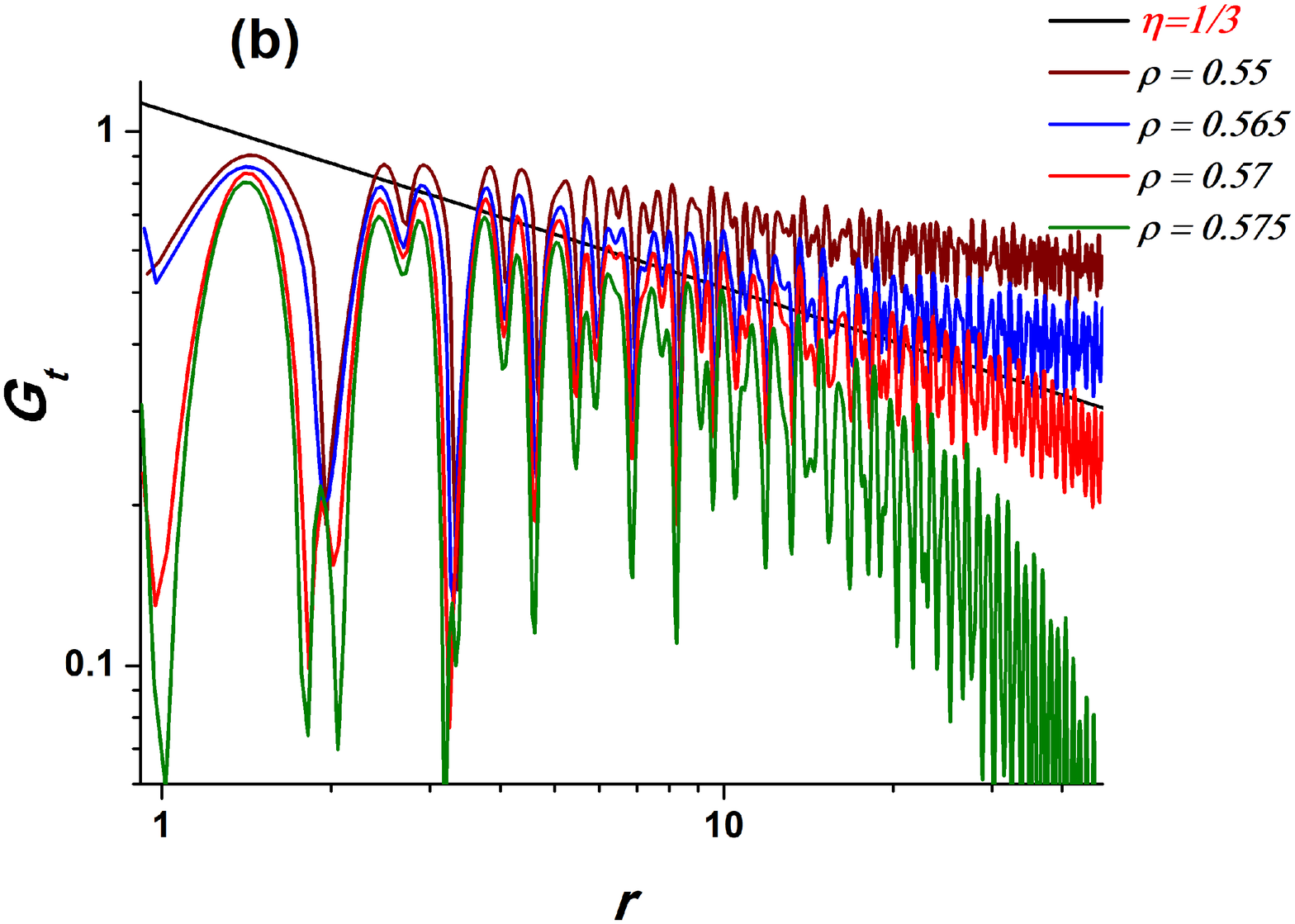}%

\caption{\label{g2-ld} (a) The orientational correlation functions
of the system with $H=1.0$ at the right branch of the melting line
of the low density triangular crystal at $T=0.1$. (b) The same for
the translational correlation functions.}
\end{figure}

Figs. \ref{g2-ld} (a) and (b) present the same correlation
functions at the right branch of the melting line. From comparison
of the stability limits obtained from the correlation functions
with the position of the Mayer-Wood loop we conclude that at the
right branch of the low density phase diagram melting occurs in
accordance with the third scenario because the crystal's stability
limit goes beyond the Mayer-Wood loop existence region.

\begin{figure}
\includegraphics[width=8cm]{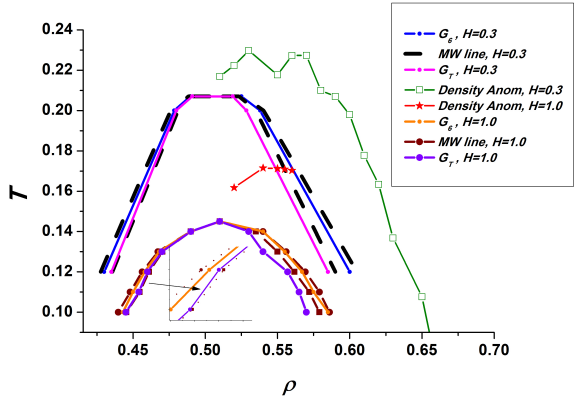}%

\caption{\label{bump} The phase diagrams of the system with
$H=0.3$ and $H=1.0$ in the vicinity of the low density triangular
crystal. The inset enlarges the part of the left branch of the
melting line at $H=1.0$. It enables us to see that a single first
order transition preempts the appearance of the hexatic phase in
the system. The lines of density anomalies are also shown.}
\end{figure}

Fig. \ref{bump} shows the phase diagrams in the vicinity of the
low density triangular crystal for the system with $H=0.3$ and
$H=1.0$. In the case of $H=0.3$ the melting lines perfectly
coincide with those of the purely $2D$ system (not shown here)
\cite{ufn1,we5}. In the case of $H=1.0$ the qualitative behavior
of the system is the same, however, the region of stability of the
crystal is shifted to lower temperatures. Thus at a small $H$ the
system is strongly confined and out-of-plane motions are
negligible. Therefore, the phase diagram coincides with that of
the purely $2D$ system. However, at a larger $H$ even if the
distribution of density shows a high narrow peak, the out-of-plane
motion destabilizes the crystal at lower densities. It is also
interesting to note that the temperature of the density anomaly
region decreases with an increase in the pore width.

\begin{figure}
\includegraphics[width=8cm]{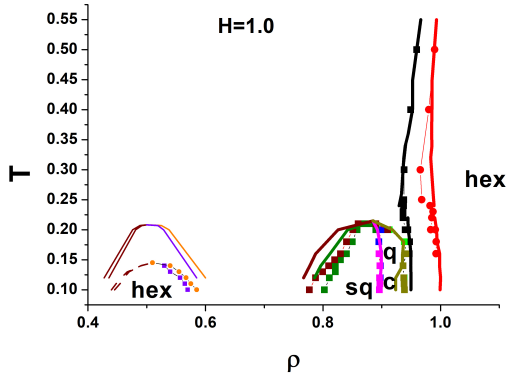}%

\caption{\label{pd} The symbols on the phase diagram of the system
with $H=1.0$ in comparison with the phase diagram of the purely
$2D$ system (solid lines).}
\end{figure}

Similar analysis is employed to study the melting lines and
coexistence regions of other phases. The phase diagram of the
system with $H=1.0$ is shown in Fig. \ref{pd}, where the phase
diagram of the purely $2D$ system is also given for comparison.
One can see that at high density the phase diagram of the confined
system is in close agreement with that of the purely $2D$ system.
Thus one can suppose that at high density the effective strength
of confinement appears to be greater, and no strong influence of
out-of-plane motion is detected.

\section{Conclusions}

In conclusion, we performed a study of the phase diagram of a
core-softened system described by the potential (\ref{pot}) Only
the case of strong confinement when a single layer of particles is
formed is considered. No changes of the phase diagram compared to
the purely $2D$ system were detected at very small slit pore width
($H=0.3$). At larger pore width ($H=1.0$) melting of the low
density triangular crystal occurred at lower temperatures compared
with the purely $2D$ case and the confined system at $H=0.3$ due
to intensive out-of-plane motions. The phase diagram at high
density remains nearly unaltered regardless of the pore width
demonstrating the negligible influence of out-of-plane motions on
the melting lines of crystalline phases. These results are in
qualitative agreement with the previous studies by other authors
\cite{rice1,rice3,rice2}. To provide a fuller picture we plan to
perform investigations of confined systems with a wider slit pore,
which leads to the appearance of two or more layers.

\bigskip

This work has been carried out using computing resources of the
federal collective usage center Complex for Simulation and Data
Processing for Mega-science Facilities at NRC "Kurchatov
Institute", http://ckp.nrcki.ru/, and supercomputers at Joint
Supercomputer Center of the Russian Academy of Sciences (JSCC
RAS). The work was supported by the Russian Science Foundation
(Grant No 19-12-00092).

\end{document}